\documentclass{pnastwo}

\usepackage[dvips]{graphicx}
\usepackage{amssymb,amsfonts,amsmath}
\usepackage[T1]{fontenc} 
\usepackage{multirow}

\usepackage{textcomp}     
\usepackage{color}

\url{http://www.pnas.org/content/109/26/10245}
\copyrightyear{2012}
\issuedate{June 26}
\volume{109}
\issuenumber{26}
\newcommand{\q}[1]{\textcolor{black}{#1}}

\newcommand{\bb}[1]{\textcolor{black}{#1}}

\begin{document}

\title{Anomalous Front Broadening During Spontaneous Imbibition in a Matrix with Elongated Pores}

\author{Simon Gruener,\affil{1}{Experimental Physics, Saarland University, D-66041 Saarbruecken, Germany}
Zeinab Sadjadi,\affil{2}{Theoretical Physics, Saarland University, D-66041 Saarbruecken, Germany}
Helen E.~Hermes,\affil{3}{Condensed Matter Physics Laboratory, Heinrich-Heine University, D-40225 Duesseldorf, Germany}
Andriy V.~Kityk,\affil{4}{Faculty of Electrical Engineering, Czestochowa University of Technology, P-42200 Czestochowa, Poland}
Klaus Knorr,\affil{1}{}
Stefan U.~Egelhaaf,\affil{3}{}
Heiko Rieger\affil{2}{}
\and
Patrick Huber\affil{1}{}\affil{5}{Department of Physics, Pontifical Catholic University, Santiago, Chile}\affil{6}{Materials Physics and Technology, Hamburg University of Technology, Hamburg, Germany}}

\contributor{Submitted to Proceedings of the National Academy of Sciences
of the United States of America}

\maketitle

\begin{article}
\begin{abstract}
\bb{During spontaneous imbibition a wetting liquid 
is drawn into a porous medium by capillary forces. In systems with comparable pore length and diameter, such as paper and sand, the front of the propagating 
liquid forms a continuous interface. Sections of this interface advance in a highly correlated manner
due to an effective surface tension, which restricts front broadening. 
Here we investigate water imbibition in a nanoporous glass (Vycor) in
which the pores are much longer than they are wide.
In this case, no continuous liquid-vapor interface with coalesced
menisci can form. Anomalously fast imbibition front roughening is
experimentally observed by neutron imaging.
We propose a theoretical pore network model, whose structural details are adapted to the microscopic pore structure of 
Vycor glass, and show that it displays the same
large scale roughening characteristics as observed in the experiment.
The model predicts that menisci movements are uncorrelated.
This indicates that despite the connectivity of the network the 
smoothening effect of surface tension on the imbibition front 
roughening is negligible. These results suggest a new universality class of imbibition behavior
which is expected to occur in any matrix with elongated,
interconnected pores of random radii.}
\end{abstract}


\abbreviations{NVG , nanoporous Vycor glass}

\dropcap{M}any everyday processes involve the flow of a liquid into a porous matrix, for instance when we dunk a biscuit into coffee, clean the floor with a cloth, or get drenched with rain. The same process is also important in nature, e.g.~for water to reach the tips of the tallest trees or to flow through soil, and crucial for different industrial processes, ranging from oil recovery and chromatography to food processing, agriculture, heterogeneous catalysis, and impregnation (for reviews see \cite{Sahimi1993, Halpinhealy1995, Hinrichsen2000, Alava04}).

\begin{figure}[b] \center
\includegraphics[width=.5\linewidth]{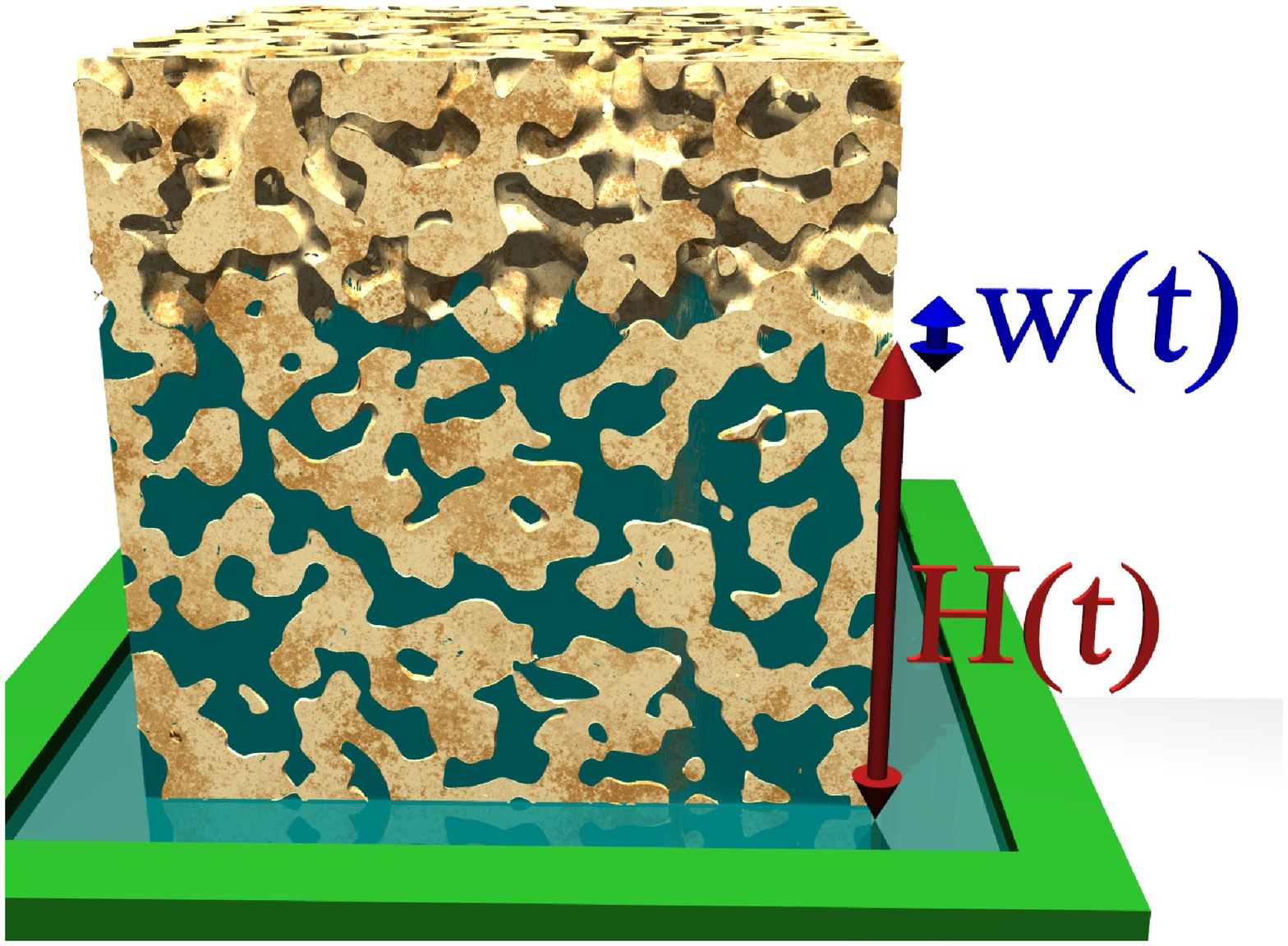}
\caption{Schematic representation of spontaneous imbibition of a fluid into a porous matrix. The arrows indicate the mean median rise level $H(t)$ and the invasion front width $w(t)$.}
\label{fig1}
\end{figure}

The above processes are examples of imbibition (Fig.~\ref{fig1}). Imbibition of a liquid
into a porous matrix is governed by the interplay of capillary
pressure, viscous drag, volume conservation and gravity. The porous
matrix often has a complex topology. The inhomogeneities result in
variations in the local bulk hydraulic permeability and in the
capillary pressure at the moving interface. Nevertheless, the invasion
front during solely capillarity-driven (i.e.~spontaneous)
imbibition advances in a simple square-root-of-time manner, according
to the Lucas-Washburn law \cite{Lucas18, Washburn21}. This is a result
of the time-independent mean capillary pressure and the increasing
viscous drag in the liquid column behind the advancing front. It is
valid down to nanoscopic pore sizes \cite{Dimitrov07, Gruener09,
Gruener09a} and particularly robust with regard to the geometrical
complexity of the porous matrix \cite{Sahimi1993, Alava04, Courbin07,
Reyssat08}. The evolution of the invasion front displays universal
scaling features on large length and time scales, which are
independent of the microscopic details of the fluid and matrix
\cite{Planet07, Dube07, Buldyrev92, Horvath95, Miranda2010,
Hernandez01, Herminghaus02}. This parallels the elegance of critical
phenomena.

Typically imbibition is studied using paper \cite{Buldyrev92,
Horvath95, Miranda2010} or Hele-Shaw cells \cite{Hernandez01,
Herminghaus02}. In these systems, pore space is laterally highly
interconnected. This results in a continuous liquid-gas interface,
whose advancement is spatially correlated due to an effective 
surface tension \cite{Dube00a}.
\bb{Consequently menisci advancement beyond the average front position 
is slowed down whilst menisci lagging behind are drawn forward. Hence, the
roughening of the interface is slowed down.} By contrast, in many real 
porous systems, e.g.~rock and soil \cite{Song00}, the pore 
network consists of elongated pores with reduced connectivity 
\cite{Sahimi1993}.

Here we investigate the spontaneous imbibition of water into
nanoporous Vycor glass (NVG), which is a silica substrate with an
interconnected network of nanometer-sized, elongated pores
\cite{Gelb98}. The narrow pores lead to capillary
pressures of several hundred times atmospheric pressure. This means
that gravity would only halt capillary rise after several kilometers
and several billion years \cite{Caupin08}. Hence, with this system, we
are able to observe pure spontaneous imbibition over large length
(centimeter) and long time (hours) scales. The observation of the
advancing front is difficult since it is deeply buried inside the
matrix \cite{Callaghan91, Howle93}. Nevertheless, neutron radiography
\cite{Winkler06, Kaestner08, Strobl09} allows us to image the liquid
inside porous materials \cite{Cnudde08, Hall10}. Recent technical
improvements in neutron imaging provide the spatial and temporal
resolution (tens of $\mu$m and tens of s) to follow imbibition and to
obtain quantitative information on the morphological evolution of the
progressing interface. 

Our experiments and simulations show that the interface roughness 
as measured by the interface width $w(t)$ increases much faster than
observed previously, namely $w(t)\propto t^\beta$ with
$\beta\approx0.45$. This is close to the square-root-of-time
progression of the invasion front H(t) $\sim t^{0.5}$. The
propagation front hence comprises an almost constant fraction,
$w(t)/H(t)$, of the occupied part of the matrix, including voids and
overhangs. We find that lateral correlations of the invasion front
are short-ranged and independent of time. This means that, in the
present case, surface tension is irrelevant in a coarse
grained description of interface broadening on macroscopic length
scales. \q{Similar broadening has been observed, but for different experimental conditions, namely drainage and forced imbibition of less-wetting fluids \cite{Lenormand1990, Martys1991}}.

\section{Results and Discussion}

{\bf Quantitative Experimental Characterization of the Imbibition
Front.} We investigate imbibition in NVG which contains pores with a
mean radius $r_{\rm av} = 4$~nm, a radius polydispersity of 20\,\%, a
pore aspect ratio $a=L/2 r_{\rm av}$ between 5 and 7 (where $L$ is the
pore length), and a porosity of about 30\,\%
\cite{Gruener09, Gelb98, Gruener10a, Levitz91}. (For details see Methods.) 
The bottom face of an empty NVG is brought into contact with the
surface of a water reservoir. Capillary forces draw the liquid 
into the porous matrix.

\begin{figure} \center
\includegraphics[width=0.5\linewidth]{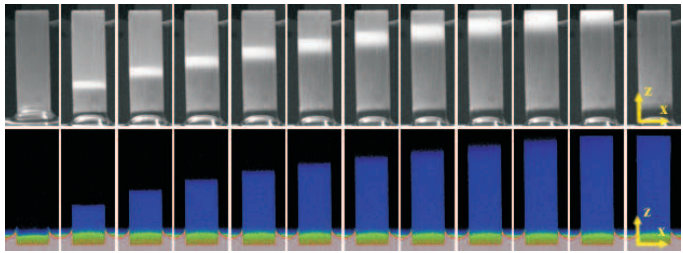}
\caption{Direct observation of spontaneous imbibition of water into nanoporous Vycor glass using visible light (top row) and neutrons (lower row). The reflected light intensity and the local liquid concentration, i.e.~the filling degree $f(x,z,t)$, are shown in grayscale and pseudocolors, respectively. Snapshots are recorded for 0.1~s (light) and 30~s (neutrons) about every 15~min. The lateral direction $x$ and the height $z$, i.e.~the direction of capillary rise, are indicated. The width and height of the sample are 4.6~mm and 20~mm, respectively.} 
\label{fig2}
\end{figure}

The invasion front appears as a bright region when imaged using
reflected light (Fig.~\ref{fig2}, top row). \q{The pronounced (multiple)
light scattering is caused by filled and empty regions randomly
alternating on the length scale of the wavelength of visible light; several hundred nanometers  \cite{Page93}. These spatial scales are much larger than any intrinsic length of the porous matrix, such as the pore diameter or the pore-pore distance. Thus strong scattering of light at the advancing interface indicates structures of filled and empty parts (voids) of surprisingly large size.} It prevents also a quantitative determination of the liquid
profile in the propagating front and hence of the width of the
interface. By contrast the spatial and
temporal resolutions of neutron imaging \cite{Lehmann07} allow
quantitative measurements in systems such as NVG
(Fig.~\ref{fig2}, bottom row).

\begin{figure} \center
\includegraphics[width=0.4\linewidth]{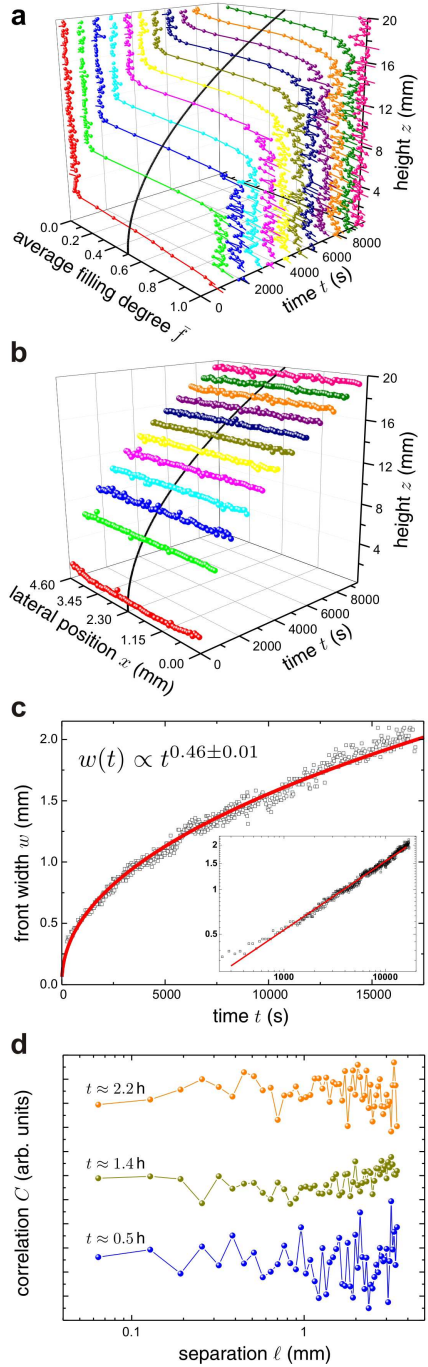}
\caption{Progression and broadening of the imbibition front of water in nanoporous Vycor glass as quantified by neutron imaging. {\bf (a)} Laterally-averaged filling degree $\bar{f}(z,t) \equiv \langle f(x,z,t) \rangle_{\rm x}$ as a function of height $z$ and time $t$. {\bf (b)} Local median rise level $h=z(f{=}0.5,x,t)$ as a function of lateral position $x$ and time $t$. {\bf (c)} Evolution of the front width $w(t)$ along with a fit of $w\propto t^\beta$ (solid line). The first few data points show an apparently increased width due to smearing effects and are thus disregarded in the fit. (See text for details.) The inset shows the same data in a log-log representation. \q{Axes units of the inset agree with the ones of the main plot.} {\bf (d)} Height-height correlation function $C(\ell,t)$ of the invasion front at three different times (as indicated). The data are shifted for clarity. In (a) and (b) the Lucas-Washburn law $z{\propto} \sqrt{t}$ is shown as solid lines. Parts (a) and (b) show only about the first half of the collected data, during which most of the front movement occurs.}
\label{fig3}
\end{figure}

From the neutron images we determine the spatial and temporal
evolution of the local filling degree $0 \le f(x,z,t) \le 1$. Due to
the projection in the $y$-direction, this is the average amount of
filled pore space at lateral position $x$, height $z$ and time
$t$. Its lateral average, that is the vertical concentration profile
$\bar{f} (z,t) \equiv \langle f(x,z,t) \rangle_{\rm x}$, is shown in
Fig.~\ref{fig3}a. The time-dependence of the front height, quantified
by the mean median rise level $H(t) \equiv \langle z(f{=}0.5,x,t)
\rangle_{\rm x}$, follows the Lucas-Washburn $\sqrt{t}$-law
(Figs.~\ref{fig3}a,b, solid lines), consistent with previous studies
\cite{Gruener09}. Fits of Gauss error functions to the profiles yield the time-dependence of the width $w(t)$ (Fig.~\ref{fig3}c). The fit of $w(t)\propto t^\beta$ results in a growth exponent of the width or
roughness, $\beta = 0.46 \pm 0.01$ (Fig.~\ref{fig3}c, solid
line). The value $\beta = 0.46$ significantly exceeds previous
theoretical predictions, in particular those from phase-field models
which are based on quenched, random fields. Such models predict slower
roughening dynamics with $\beta \approx 0.19$ and a strong spatial
correlation of the height fluctuations within the moving interface
\cite{Dube07}.

Instead of the median rise level averaged in $x$ direction, $H(t)= \langle z(f{=}0.5,x,t)\rangle_x$, we now consider the local $x$-dependent median rise level $h(x,t) \equiv z(f{=}0.5,x,t)$ (Fig.~\ref{fig3}b) to investigate fluctuations in $x$-direction, i.e. within the front. We calculate the height-height correlation function:
\begin{equation}
C(\ell,t) = \sqrt{ \langle (h(x,t)-h(x{+}\ell,t))^{\rm 2}  \rangle_{\rm x} }\;.
\label{eq:corrfunction}
\end{equation}
The observed fluctuations in $C(\ell,t)$ (Fig.~\ref{fig3}d) are due mainly to the limited data density and stray gamma radiation from
the reactor and instrument hitting the camera. The data exhibit
neither scaling of $C(\ell,t)$ with $\ell$ nor any indication of
spatial correlations in the experimentally accessible range 75\,\textmu m~$\le \ell\le$~4000\,\textmu m. Although the correlations 
are reduced due to the projection in $y$-direction, the absence of any
detectable correlation is in contrast to all previously reported
experiments and theories on imbibition front roughening.

\bigskip

{\bf Pore-network Model.} 

No theoretical model is available that is consistent with our
system and which predicts the spontaneous imbibition behavior observed.
\bb{An ensemble of independent pores of random but constant radius
exhibits a roughening exponent 1/2 since the meniscus heights evolve
independently from one another as $h_i(t)=a_i \sqrt{t}$ 
with random pre-factors $a_i$. However, this independent pore model is
inappropriate for Vycor glass, since the pore radii vary strongly
along individual pores (see Fig. 1).} An ensemble of independent pores
with radii which vary randomly along their length has a roughening
exponent of 1/4 (see Appendix), which does not agree with the experimentally
observed value. Thus independent pore models do not explain the
observed exponent. \bb{A roughening exponent $\beta \approx
1/2$ has recently been reported within the framework of a lattice gas
model for spontaneous imbibition \cite{Leoni11}. This model is
appropriate for silica aerogels with an extremely large porosity of
87{--}95\,\% and gives rise to a continuous liquid-gas interface. 
Consequently one expects here an effective 
surface tension to be present, inducing height-height correlations
in the advancing imbibition front. The model details are thus not appropriate for NVB.}
To our knowledge, all other existing theoretical models (for an overview
see \cite{Alava04}) are also incompatible with our experimental observations
which are i) fast broadening dynamics with a growth exponent close to
$1/2$, ii) absence of height-height correlations in the advancing
imbibition front.

Hence we propose a pore-network model \cite{Aker98,
Lam00} adapted to our experimental situation. This consists of
individual, elongated capillaries arranged in a two-dimensional square
lattice with laterally periodic
boundary conditions. Capillaries are connected at nodes and inclined at $45^\circ$. All
capillaries have the same length $L$, while the radius $r$ of each
capillary is randomly chosen from a uniform distribution with mean
radius $r_{\rm av}$ and width $2\delta_{\rm r}$, i.e.~disorder
strength $\delta_{\rm r}/r_{\rm av}$.
\cite{Gelb98, Levitz91}. (For details see Methods.)
We investigated aspect ratios 
$2.5 \le a \le 10$ and polydispersities 
$0.1 \le \delta_{\rm r}/r_{\rm av} \le 0.4$.

The pressure at all nodes at the bottom of the lattice are set to zero
while at the menisci the Laplace pressure prevails. This pressure
difference drives the flow through the capillaries. This flow is
opposed by viscous drag according to Hagen-Poiseuille's law. During
the whole process, volume conservation must be maintained.
\bb{When a meniscus reaches an empty node it `jumps' over the
node, generating new menisci in a distance $\delta = L/100$ from the
node (Fig.~\ref{fig4}a). This implementation of node crossing avoids a
microscopic treatment of the filling process of the nodes and is valid
as long as this is not the rate limiting step, which we will discuss below.
If the Laplace pressure of a meniscus exceeds the node pressure the
meniscus is arrested at the distance $\delta$ until the node
pressure increases beyond the Laplace pressure
(Fig.~\ref{fig4}b). Then propagation of the meniscus resumes.}

\bigskip

{\bf Computer Simulation Results of Imbibition.} With our model, we observe a strong roughening of the imbibition front (Fig.~\ref{fig4}d) with fast moving menisci advancing through
sequences of thin capillaries and arrested menisci lagging behind. Quantitatively, the computer simulations yield a
mean rise level $H(t)=\langle h_i(t)\rangle$ (where
$\langle\ldots\rangle$ denotes an average over all menisci labelled by
the index $i$), which obeys the Lucas-Washburn $\sqrt{t}$-behavior
(Fig.~\ref{fig5}a). The width $w(t)=(\langle h_i^2(t)\rangle-\langle
h_i(t)\rangle^2)^{1/2}$ increases rapidly as
$w(t)\propto t^\beta$ with $\beta=0.42\pm0.01$ or $0.45\pm0.01$ for
the smallest and largest polydispersities, respectively (Fig.~\ref{fig5}b). 
We find a slight upward trend of $w(t)$ at large times
(Fig.~\ref{fig5}b, inset), which is also suggested in the experimental
data (Fig.~\ref{fig3}c) and indicates that the asymptotic value of the growth exponent might be larger. This is consistent with the fact that a
smaller $\beta$ is observed when the asymptotic behavior is
approached later (as in the case of the smaller polydispersity). This
implies that the asymptotic value is closer to that found for the
larger polydispersity with an increased uncertainty, i.e.\
$\beta=0.45\pm0.02$. Variation of the
aspect ratio in the range $2.5\le a \le 10$ gave identical results for
the exponent $\beta$.

We systematically studied finite size effects,
especially on $w(t)$. Remarkably, we only find a dependence on the
lateral system size $N_{\rm x}$ for the smallest system size $N_{\rm
x}=4$ (Fig.~\ref{fig5}c). This provides an upper bound for the
characteristic length scale $\xi(t)$ of height-height
correlations. Within the framework of the scaling theory of roughening
\cite{Barabasi95, Krug97}, the interface width in a finite system of
lateral size $N_x$ is expected to behave as
\begin{eqnarray}
w_{N_x}(t) 
& \sim & 
\left\{
\begin{array}{lcl}
t^{\beta}        &           {\rm for }   &   \xi(t)/L \ll N_x\\
{\rm const.}   &           {\rm for }   &   \xi(t)/L \gg N_x\\
\end{array}
\right.
\end{eqnarray}
The data (Fig.~\ref{fig5}c) suggest $\xi(t)<4L$, implying that the
roughening dynamics are not or are only weakly spatially
correlated. This is confirmed by the height-height correlation
function $C(\ell,t)$ (eq.~\ref{eq:corrfunction}, Fig.~\ref{fig5}d),
which saturates quickly (around $\ell\approx L$). Scaling theory
\cite{Barabasi95, Krug97} predicts saturation of $C(\ell,t)$ for
$\ell\gg\xi(t)$, which implies $\xi(t)/L={\cal O}(1)$ independent of
time $t$. This finding supports our experimental result
(Fig.~\ref{fig3}d) that any spatial roughness correlations are absent
and extends its validity down to pore-pore distances and thus towards
the nanometer-scale, i.e.~far beyond our experimental resolution.

\bigskip

{\bf Experiments and Simulations Both Yield an Anomalous Roughness
Growth Exponent.} Experiments and simulations exhibit corresponding
behavior, even on a quantitative level, i.e.~progression of the
imbibition front according to the Lucas-Washburn $\sqrt{t}$-law, fast
broadening of the front with a large growth exponent $\beta \approx
0.45$ and short range height-height correlations over a maximum of
1--2 pore lengths only. The pore-network model with its elongated
pores hence successfully mimics the characteristics of NVG. 
\q{In such morphologies, all menisci are restricted to individual pores and
thus cannot interact via an effective surface tension. 
In the absence of interactions between individual menisci local
processes at the junction become important for the front
roughening. The dynamics of junction filling, as analyzed in
\cite{Shikhmurzaev12}, describes a threshold mode in which meniscus
propagation is halted whilst the junction is filled and
the new menisci in the adjacent pores form. This takes only a few 
milliseconds in nanometer sized pores with filling heights up to 2 cm (as in the present case), for which gravity is negligible. Once the new menisci have formed, those in the thicker pores 
are arrested as long as their Laplace pressure is larger than the 
pressure within the junction. These arrests can last much longer than the filling process, up to times of the order of the age of the 
propagation front, which can be several hours in our experiments. Thus for the asymptotic (long time) behavior of the broadening
dynamics, the filling process of individual junctions is negligible
and front roughening is mainly influenced by the arrests after
the filling of the junctions. As a consequence, the distribution of 
pore diameters and the frequencies of junctions is expected to
be much more important than the topology, in particular the 
dimensionality, of the network. This contrasts with the role 
the latter usually plays in surface roughening and critical phenomena 
\cite{Barabasi95, Krug97}.}

It should be noted that it is crucial that the pores in NVG are
connected \cite{Gelb98,Levitz91}, since an ensemble of
independent pores with randomly varying radii over their lengths would
give a roughness exponent 1/4 (see Appendix). Perhaps
counterintuitively, the introduction of junctions, i.e.~branch points
or crossings, enhances height fluctuations and hence the front
width. This is because at each branch point one meniscus can split
into two (or more) upwards moving menisci with one typically moving
faster than the other, or even one moving and the other stopping until the node pressure exceeds its Laplace pressure.

\section{Conclusions}

{\bf Pore Aspect Ratio Determines the Universality Class.} Our results
show that spontaneous imbibition crucially depends on the pore aspect
ratio $a$.  For short pores (small $a$), neighboring menisci
coalesce and form a continuous imbibition front. Thus the smoothening
effect of an effective surface tension within the interface leads to a
slow broadening of the front. Various theoretical models
describe the roughening of a vapor-liquid interface during spontaneous
imbibition in the presence of an effective surface tension, e.g.\ phase field
models \cite{Dube00a}. These predict a roughening exponent $\beta
\approx 0.19$. \q{The elongated pores in nanoporous Vycor glass 
(large aspect ratio $a$) inhibit the formation of a
connected vapor-liquid interface. In this case the individual menisci cannot interact via an effective surface tension} and the
broadening of the imbibition front is anomalously fast with $\beta
\approx 1/2$ establishing another universality class.
\q{The regime of weak roughening (small $a$) must be separated from
the regime of strong roughening (large $a$) by a critical value $a_c$ 
of the aspect ratio, but its precise
value will depend on structural details of the pore
network, in particular the pore junction geometry.}

We want to stress that strong imbibition front broadening is not
linked to the nanometer size of the pores. However, its experimental
observation over large length and time scales significantly benefits
from the dominance of capillary forces over gravitational forces,
which results from the nanometer-sized pores. The theoretical model
employs macroscopic hydrodynamic concepts only. Therefore, strong
interfacial broadening is a consequence of any spontaneous imbibition
process in porous structures with interconnected elongated capillaries
independent of their macroscopic extension and mean pore diameter. 
It is not only important to nanofluidics,
but for liquid transport in porous media in general. 

Our observation of a new universality class of strong interfacial
broadening is thus a very general finding, which has been made
possible due to recent improvements in the resolution of neutron
imaging \cite{Lehmann07}. The front
roughness is crucial for many processes, such as water transport in
geology, flux in oil recovery, glueing, dying and
impregnation. Our results enable us to link the broadening dynamics
during these processes to the properties of the porous materials. To
what extent this behavior can be described with alternative models for
transport in porous media, e.g. models which consider a
saturation-dependent hydraulic permeability of the pores
\cite{Alava04}, warrants further investigation.

\begin{figure} \center
\includegraphics[width=0.5\linewidth,angle=0]{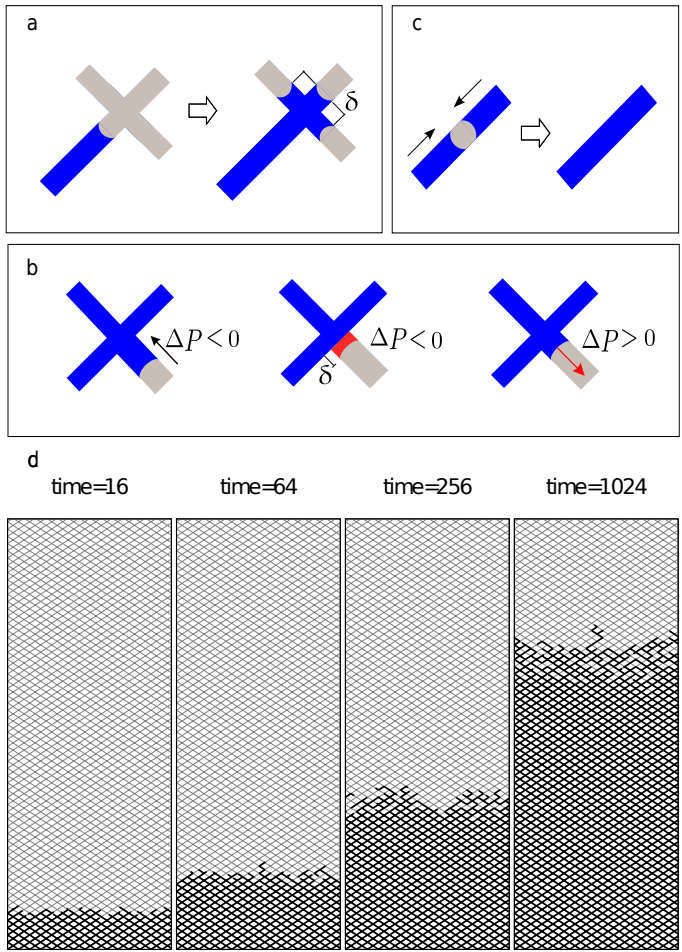}
\caption{Computer simulation of spontaneous imbibition in a 
network of elongated capillaries. Rules for the propagation of
menisci are designed to mimic the experimental situation. {\bf (a)}
After reaching an empty node, the liquid immediately fills the
connected capillaries for a distance $\delta$. {\bf (b)} After
retracting up to $\delta$ toward a filled node, the meniscus is
arrested until the pressure difference driving the liquid is again
positive. {\bf (c)} When two menisci meet, they merge. {\bf (d)}
Snapshots of configurations in a system with aspect ratio $a=5$, 
radius polydispersity
$\delta_{\mathrm{r}}/r_{\mathrm{av}}=0.3$ and lattice size $16 \times 64$ at
different times (as indicated in units of ns).}
\label{fig4}
\end{figure}

\begin{figure} \center
\includegraphics[width=0.4\linewidth]{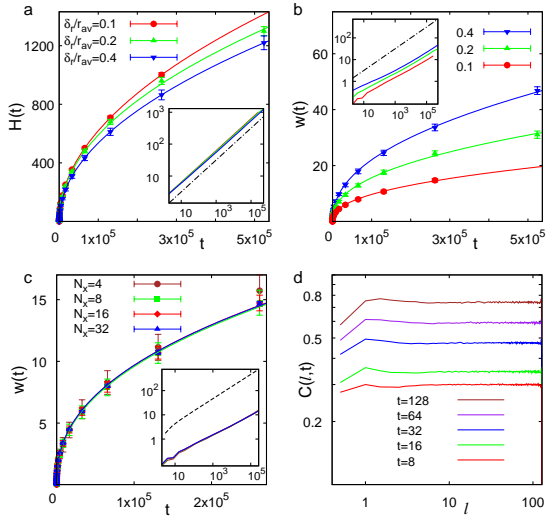}
\caption{Progression and broadening of the imbibition front as 
observed in computer simulations based on our pore-network model. 
Pore aspect ratio $a=5$ and standard lattice size
$N_{\rm x} \times N_{\rm z}=16 \times 1000$. Lengths are measured in
units of $L$, times in units of ns. All data are averaged over 100
simulations with different disorder realizations, the error bars
reflect the standard deviation.
{\bf (a)} Mean median rise level $H(t)=\langle h_i(t)\rangle_i$ for
different polydispersities $\delta_{\rm r}/r_{\rm av}$ (as
indicated). The lines represent fits of $H\propto \sqrt{t}$. The inset
shows the same data in a log-log representation and $H \propto
\sqrt{t}$ as the dash-dotted line. 
{\bf (b)} Evolution of the front width $w(t)$ for different
polydispersities (as indicated). The lines
represent fits of $w(t) \propto t^{\beta}$ with growth exponents
$\beta = 0.42\pm 0.01$, $0.42\pm 0.01$ and $0.45\pm 0.01$ for
polydispersities of $0.1$, $0.2$ and $0.4$,
respectively. The inset shows the same data in a log-log
representation and $\sqrt{t}$ as the dash-dotted line. 
{\bf (c)} Evolution of the front width $w(t)$ for a polydispersity
$\delta_{\rm r}/r_{\rm av}=0.1$ and different lateral system sizes
$N_{\rm x}$ (as indicated). The line represents a fit of $w
\propto t^\beta$ with $\beta=0.42\pm0.01$ for $N_{\rm x} > 4$. The
inset shows the same data in a log-log representation, the dashed line
represents an upper bound of the width, which is given by the
difference between the front heights $H(t)\propto \sqrt{t}$ in two
homogeneous systems with constant minimal and maximal capillary radius
$r=r_{\rm av}{-}\delta_{\rm r}$ (slowest front propagation) and
$r=r_{\rm av}{+}\delta_{\rm r}$ (fastest front propagation),
respectively. 
{\bf (d)} Height-height correlation $C(\ell,t)$ as a
function of the distance $\ell$ at different times $t$ (as indicated)
for a sample with polydispersity $\delta_{\mathrm{r}}/r_{\mathrm{av}}=0.1$ and
$N_{\rm x}\times N_{\rm z}=128 \times 32$.}
\label{fig5}
\end{figure}

\begin{materials}

{\bf Neutron Imaging.} 

The nanoporous Vycor glass (NVG) consists of an
interconnected network of elongated pores with a mean radius 
$r_{\rm av} = 4$~nm, a radius polydispersity
$\delta_{\mathrm{r}}/r_{\mathrm{av}}=0.2$, a pore aspect ratio 
$5\lesssim a \lesssim 7$, and a porosity of about 30\,\%
\cite{Gruener09, Gelb98, Gruener10a, Levitz91}. The macroscopic
dimensions of the sample are $4.6 \times 4.6 \times 20$\,mm$^{\rm 3}$. 
Its faces, except the bottom face, are sealed to preclude liquid
evaporation. To initiate imbibition, the bottom face of the sample is
brought into contact with the surface of a water reservoir. During
imbibition, the huge capillary pressure highly compresses entrapped
air which is subsequently dissolved in water and hence does not affect
our experiments. All experiments are performed at room temperature.

The neutron imaging experiments are performed at the ANTARES beamline
of the research reactor FRM~II of the Technical University Muenchen
(Garching, Germany) \cite{Calzada09}. A beam of cold neutrons passes
through an aperture with size $D$ and, after a distance $L$,
`illuminates' the sample, which is situated $d=30$~mm in front of the
scintillator. The geometrical resolution is $d/(L/D)=75\;\mu$m. The
transmitted neutrons are detected using a very thin `Gadox'
scintillator, which does not limit the geometrical resolution, and a
CCD camera with pixel size $15.97$\,\textmu m. Series of images are
recorded for total measurement times up to several hours.
Individual measurement times are 30~s and data transfer
times 10~s.  In the
first few kinetic images smearing occurs due to the front moving a
significant distance during the individual measurement times. However,
after about 1000~s, the smearing due to the limited time resolution is
negligible compared to the spatial resolution (and only these data are 
used for fitting $w(t)$). Raw images were
corrected for detection efficiency, background and noise, while
corrections for scattered neutrons are not necessary
\cite{Hassanein05}.

The experimentally determined neutron transmission $T(x,z,t)$, that is
the ratio of transmitted intensity $I(x,z,t)$ and incident intensity
$I_0(x,z,t)$, is related to the absorption coefficient $S(x,z,t)$ by
\begin{equation}
T(x,z,t) = \frac{I(x,z,t)}{I_0(x,z,t)} = e^{-S(x,z,t)\, d}
\end{equation}
where $d=4.6$~mm is the sample thickness. The absorption coefficient
\begin{equation}
S(x,z,t) = S_{\mathrm{m}}(x,z,t) + f(x,z,t) \, S_{\mathrm{w}}
\end{equation}
depends on the absorption coefficient of the porous matrix
$S_{\mathrm{m}}(x,z,t)$, experimentally determined from the dry
matrix, and on that of the liquid $S_{\mathrm{w}}$, determined from
the completely filled matrix providing
$S_{\mathrm{m}}(x,z,t){+}S_{\mathrm{w}}$. The filling factor
$f(x,z,t)$ can then be determined from the experimentally determined
transmission $T(x,z,t)$. While silica, and thus NVG, is almost
transparent to neutrons, the neutron beam is strongly attenuated by
hydrogen in the water. The contrast is further enhanced by the
characteristic wavelength distribution of the ANTARES beamline, which
contains a large fraction of cold neutrons.

{\bf Computer Simulations.} The pore-network model consists of
capillaries arranged on a two-dimensional square lattice inclined at
$45^{\circ}$. The system consists of $N_{\rm x}$ and $N_{\rm z}$ nodes
in the horizontal and vertical directions,
respectively, with periodic boundary conditions in horizontal
direction. At the nodes, four capillaries are connected to each other
(Fig.~\ref{fig4}). All capillaries have the same length $L$, while the
radius of each capillary is chosen randomly from a uniform
distribution with mean radius $r_{\rm av}$ and distribution width
$\delta_{\rm r}$, i.e.~disorder strength $\delta_{\rm r}/r_{\rm
av}$. We performed computer simulations for different lateral system
sizes $4 \le N_{\rm x} \le 32$ and a vertical size up to $N_{\rm z}=1000$, 
which implies a maximum height $H=\sqrt{2}\,L\,N_{\rm z}$ which
was not reached by the invasion front within the simulation time. 

The water rises spontaneously from the bottom to the top of the
lattice. The dynamics are controlled by capillary pressure, viscous drag and
volume conservation. At each meniscus, i.e.~for each capillary $j$
connected to node $i$, we calculate the capillary pressure given by
the Laplace pressure
\begin{equation}
P_{{\rm c,}i}^{j} = \frac{2 \sigma}{r_i^j} \;\; ,
\end{equation}
where $r_i^j$ is the radius of the capillary and $\sigma$ the surface
tension ($\sigma=72$~mN/m for water). Flow through the capillary is
driven by the pressure difference $\Delta P_i^j = P_i - P_{\rm
c{,}i}^j$, where $P_i$ is the pressure at node $i$.

According to Hagen-Poiseuille's law, the volume flux $Q_i^j$ from node 
$i$ into capillary $j$ is
\begin{equation}
Q_i^j = - \frac{\pi (r_i^j)^4}{8 \eta} \; \frac{\Delta P_i^j}{h_i^j} \;\; ,
\end{equation}
where $h_i^j$ is the length of the liquid column in capillary $j$ of
node $i$ and $\eta$ the viscosity of the liquid ($\eta=0.88$~mPa~s for
water). The volume flux $Q_i^j$ determines the change of the liquid
volume $V_i^j$ and thus of the length of the liquid column $h_i^j$
according to $Q_i^j = {\rm d}V_i^j/{\rm d}t = \pi (r_i^j)^2\,{\rm
d}h_i^j / {\rm d}t$. Hence, once the node pressures $P_i$ are known,
the time dependencies of the heights $h_i^j(t)$ are given by ordinary
differential equations.

The node pressures $P_i$ are determined by the boundary conditions and
volume conservation. The boundary conditions are the Laplace pressure
at the menisci, $P_{{\rm c,}i}^j$, and zero pressure at all nodes at
the bottom of the lattice which are connected to the water
reservoir. The volume conservation at each node is given by
\begin{equation}
\sum_j Q_i^j=0  \;\;, 
\end{equation}
which corresponds to Kirchhoff's law. The sum runs over all
capillaries $j$ attached to node $i$. The resulting set of sparse
linear equations is numerically solved to obtain the node pressures
$P_i$ for a given meniscus height configuration $h_i^j$. The
differential equations for $h_i^j$ are then numerically integrated
using an implicit Euler scheme for time-stepping. Note that, due to
the nanometer-sized capillaries, capillary pressure dominates gravity,
which can thus be neglected.

\q{The time step $\Delta t$ in the numerical integration of the 
equations of motion of the menisci heights is chosen such that each
meniscus moves at most a distance $L/10$ and no meniscus crosses a
node. If this would occur for one meniscus, $\Delta t$ is reduced such
that this meniscus reaches the next node and then `jumps' over the
node, generating new menisci in a distance $\delta = L/100$ from the
node (Fig.~\ref{fig4}a), and all other menisci are also processed with
the reduced $\Delta t$.} 
Similarly, if the meniscus retracts due to a negative pressure
difference, $\Delta P_i^j < 0$, the meniscus is arrested when it has
approached the node up to a distance $\delta$. Thus a liquid column
with a length of at least $\delta$ is kept in the capillary,
i.e.~$h_i^j \ge \delta$ always holds. The meniscus is released when
$\Delta P_i^j > 0$ (Fig.~\ref{fig4}b). When two menisci meet, they
merge and the capillary thus is completely filled (Fig.~\ref{fig4}c),
which mimics the absence of entrapped air in our experimental
system.

During a computer simulation of the time evolution of the model, the
average rise level $H(t)=\langle h_i(t)\rangle_i$ of the invading
front and its width $w(t)=(\langle h_i^2(t)\rangle_i-\langle
h_i(t)\rangle_i^2)^{1/2}$ are calculated at different times $t$. Since
the invasion front contains overhangs and voids, the average
$\langle\ldots\rangle_i$ is taken over all menisci indexed by $i$. The
presented data are averaged over 100 simulation runs using different
disorder realizations. The statistical error of this average is
represented by the error bars of the simulation results.

\end{materials}

\appendix[Imbibition front broadening in an inhomogeneous porous medium 
of independent pores] For comparison with the proposed model, we
here consider spontaneous imbibition in an ensemble of independent,
i.e.~non-connected or isolated, pores. The radius of a single pore
varies randomly with height $h$ such that an appropriate model for the
meniscus motion in such a pore is $dh/dt=\kappa(h)/h$, where
$\kappa(t)$ is uncorrelated white noise with mean $c$ and variance
$\sigma$, i.e.~$\langle\kappa(h)\rangle=c$,
$\langle\langle\kappa(h)\kappa(h')\rangle\rangle=\sigma\,\delta(h-h')$. For
the time $T$ to reach some height $H$ one thus gets $T(H)=\int_0^H
dh\,h\,\xi(h)$ where $\xi(h)$ is white noise with mean $1/c$ and
variance $\sigma'$.  Averaging the stochastic variable $T$ yields
$\langle T\rangle\propto H^2$ (which is the Lucas-Washburn law) and
for the variance $\Delta T^2=
\langle (T-\langle T\rangle)^2\rangle\propto H^3$, 
which means $\Delta T\propto T^{3/4}$. The time to reach 
height $H$ therefore varies typically between $H^2+\Delta T$ and
$H^2-\Delta T$, vice versa at time $T$ one then expects the height
$H(T)$ to vary between $(T-\Delta T)^{1/2}$ and $(T+\Delta T)^{1/2}$
which means $\Delta H\approx \Delta T/T^{1/2}=T^{1/4}$.\\

S.G. an Z.S. contributed equally to this work. S.G. and H.H. carried out the neutron radiography experiments, Z.S. performed the computer simulations for the pore network model. S.G., H.H., S.E. and P.H. designed, discussed and analyzed the experiments, Z.S. and H.R. designed and
analyzed the theoretical model. All authors contributed to the writing of the manuscript.\\

\begin{acknowledgments}
We acknowledge FRM~II for providing beam time. We are grateful to our local contacts Michael Schulz, Elbio Calzada and Burkhard Schillinger. We thank Mikko Alava for helpful discussions. Part of this work was supported by the DFG priority program 1164, Nano- \& Microfluidics (Grant. No. Hu 850/2) and the DFG graduate school 1276, `Structure formation and transport in complex systems' (Saarbruecken).\\
\end{acknowledgments}

\end{article}

\end{document}